\documentclass{PoS}
\usepackage{cite}

\title{N$^3$LO approximate results for top-quark differential cross sections and forward-backward asymmetry}

\ShortTitle{N$^3$LO approximate results for top-quark  differential cross sections ...}

\author{\speaker{Nikolaos Kidonakis}\thanks{This material is based upon work supported by the National Science Foundation under Grant No. PHY 1212472.}\\
    Department of Physics, Kennesaw State University, Kennesaw, GA 30144, USA\\
        E-mail: \email{nkidonak@kennesaw.edu}}

\abstract{I present a calculation of approximate N$^3$LO corrections from NNLL soft-gluon resummation for differential distributions in top-antitop pair production in hadronic collisions. Soft-gluon corrections are the dominant contribution to top-quark production and closely approximate exact results through NNLO. I show aN$^3$LO results for the total $t{\bar t}$ cross section, the top-quark $p_T$ and rapidity distributions, and the top-quark forward-backward asymmetry. The higher-order corrections are significant and they reduce theoretical uncertainties.}

\FullConference{XXIII International Workshop on Deep-Inelastic Scattering\\
                 27 April - May 1 2015\\
                 Dallas, Texas} 

\def\beq{\begin{equation}}
\def\eeq{\end{equation}}
\def\beqa{\begin{eqnarray}}
\def\eeqa{\end{eqnarray}}

\begin{document}

\section{Introduction}

The calculation of higher-order corrections for $t{\bar t}$ total cross 
sections, top-quark transverse momentum ($p_ T$) and rapidity distributions, 
and the top forward-backward asymmetry ($A_{\rm FB}$) is an important part of 
top-quark physics.
QCD corrections are very significant for top-antitop pair production.
Soft-gluon corrections, calculated appropriately, are the dominant part of 
these corrections at LHC and Tevatron energies. The soft corrections are 
currently known through N$^3$LO \cite{NKaNNNLO,NKpty,NKafb}.

The soft-gluon terms in the $n$th-order perturbative corrections involve 
$[\ln^k(s_4/m_t^2)]/s_4$ 
with $k \le 2n-1$ and $s_4$ the kinematical distance from partonic threshold.
We resum these soft corrections at NNLL accuracy via factorization and renormalization-group evolution of soft-gluon functions \cite{NK2010}. 
The calculation is for the double-differential cross section 
using the standard moment-space resummation in perturbative QCD.
The first N$^3$LO expansion was given in \cite{NK2000} with a complete formal 
expression given in \cite{NK2005}.
Approximate N$^3$LO (aN$^3$LO) total and differential cross sections from the expansion of the NNLL resummed expressions have been obtained most recently in \cite{NKaNNNLO,NKpty}.
The latest aN$^3$LO results for the total cross section \cite{NKaNNNLO}, top $p_T$ and rapidity distributions \cite{NKpty}, and the top forward-backward asymmetry $A_{\rm FB}$ \cite{NKafb}, provide the best and state-of-the-art theoretical predictions.

It has been known for some time that the partonic threshold approximation in our formalism works very well for LHC and Tevatron energies; the differences between approximate and exact cross sections at both NLO and NNLO are at the per mille level. This is also true for $p_T$ and rapidity distributions and $A_{\rm FB}$. 
The use of a fixed-order expansion removes the need for a prescription to deal with divergences and the unphysical effects of such prescriptions. The 
stability and robustness of the theoretical higher-order results in our
resummation approach over the past two decades as well as the correct prediction of the size of the exact NNLO corrections validate our formalism.

\section{Top-antitop pair total cross sections at the LHC and the Tevatron}

\begin{figure}
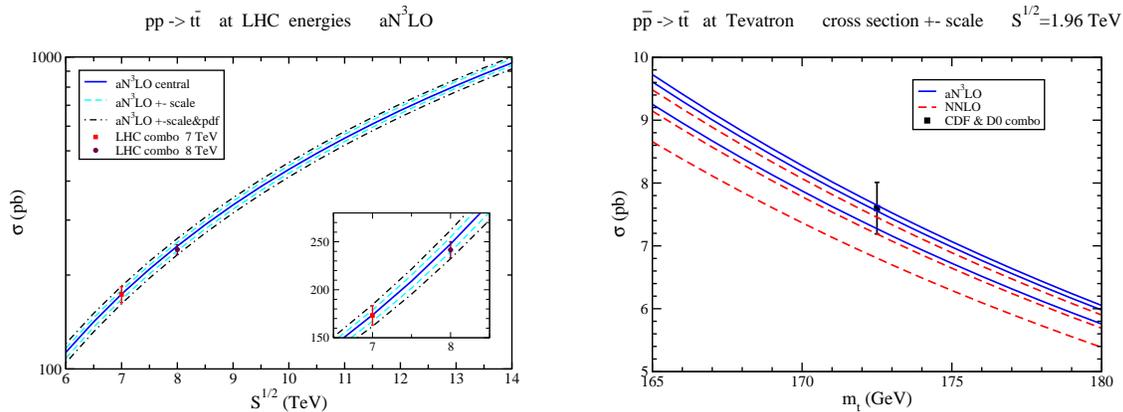

\begin{center}
\includegraphics[width=.45\textwidth]{ttSlhcaN3LOplot.eps}
\hspace{10mm}
\includegraphics[width=.45\textwidth]{tevatronaN3LOplot.eps}
\caption{Total aN$^3$LO cross sections for $t{\bar t}$ production at the LHC (left) and the Tevatron (right) and comparison with LHC \cite{ttbar7lhccombo,ttbar8lhccombo} and Tevatron \cite{ttbartevcombo} data.}
\label{figttbar}
\end{center}
\end{figure}

In Fig. \ref{figttbar} we show the aN$^3$LO total $t{\bar t}$ cross sections 
at LHC and Tevatron energies \cite{NKaNNNLO} and compare them with LHC combination data from the ATLAS and CMS collaborations at 7 TeV \cite{ttbar7lhccombo} and 8 TeV \cite{ttbar8lhccombo} energies, and Tevatron combination data from the CDF and D0 collaborations at 1.96 TeV energy \cite{ttbartevcombo}. We use MSTW2008 NNLO pdf \cite{MSTW2008} for all our predictions. The agreement of theoretical predictions with experimental data is excellent. 

We also provide the aN$^3$LO total $t{\bar t}$ cross sections with $m_t=173.3$ GeV below.
At the Tevatron with 1.96 TeV energy the cross section is $7.37 {}^{+0.09}_{-0.27} {}^{+0.38}_{-0.28}$ pb;
at the 7 TeV LHC it is $174 {}^{+5}_{-7}  {}^{+9}_{-10}$ pb;
at the 8 TeV LHC it is $248 {}^{+7}_{-8}  {}^{+12}_{-13}$ pb;
at the 13 TeV LHC it is $810 {}^{+24}_{-16}{}^{+30}_{-32}$ pb;
and at the 14 TeV LHC it is $957 {}^{+28}_{-19}{}^{+34}_{-36}$ pb. The first uncertainty in the previous numbers is from scale variation over $m_t/2 \le \mu \le 2m_t$ and the second is from the MSTW2008 pdf \cite{MSTW2008} at 90\% C.L.

Fractional contributions to the perturbative series for the $t{\bar t}$ cross section at the LHC converge well through N$^3$LO, which could potentially indicate that corrections beyond N$^3$LO are negligible \cite{NKaNNNLO}. For Tevatron energies the convergence is slower \cite{NKaNNNLO}.

\section{Top-quark $p_T$ and rapidity distributions at the LHC and the Tevatron}

\begin{figure}
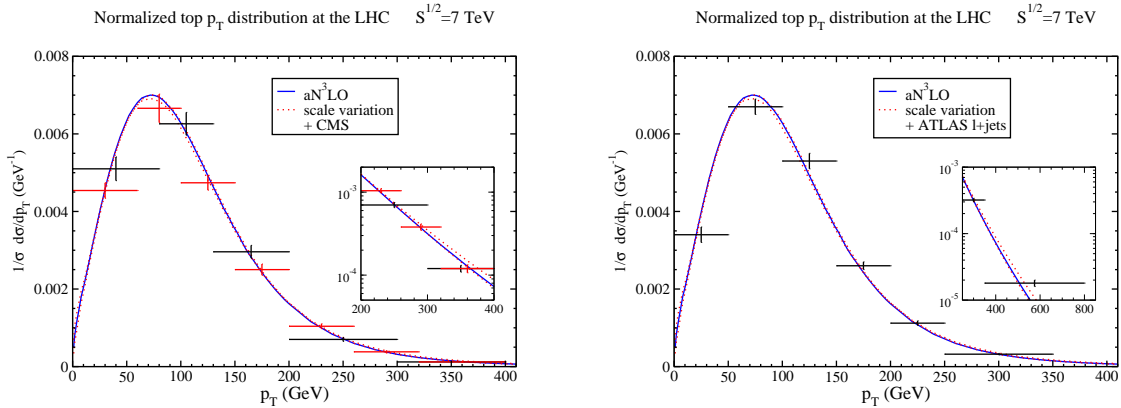

\begin{center}
\includegraphics[width=.45\textwidth]{pt7lhcnormCMSplot.eps}
\hspace{10mm}
\includegraphics[width=.45\textwidth]{pt7lhcnormATLASleptjetplot.eps}
\caption{Normalized aN$^3$LO top-quark $p_T$ distributions at the 7 TeV LHC, and comparison with CMS data \cite{CMStoppty7lhc} in the dilepton (black) and lepton+jets (red) channels (left plot), and with ATLAS data \cite{ATLAStoppt7lhc} in the lepton+jets channel (right plot).}
\label{figpt7lhc}
\end{center}
\end{figure}

In Fig. \ref{figpt7lhc} we show the normalized aN$^3$LO top-quark $p_T$ distribution, $(1/\sigma) d\sigma/dp_T$, at 7 TeV LHC energy and compare with results from CMS in the dilepton and lepton+jets channels \cite{CMStoppty7lhc} and from ATLAS in the lepton+jets channel \cite{ATLAStoppt7lhc}. We find excellent agreement between the theoretical results and the 7 TeV LHC data. The theoretical predictions are also in excellent agreement with recent CMS top $p_T$ data at 8 TeV in both channels \cite{CMStoppty8lhc}.

\begin{figure}
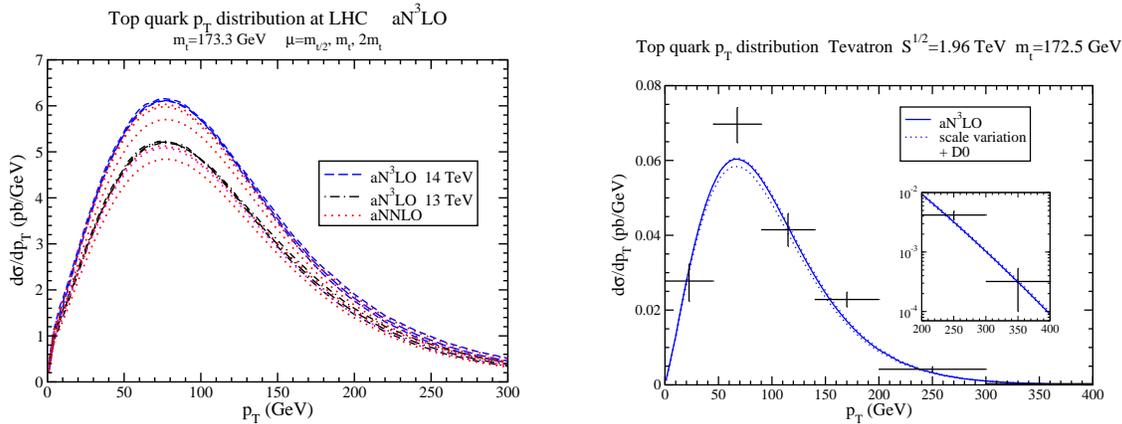

\begin{center}
\includegraphics[width=.45\textwidth]{pt13and14lhcaN3LOplot.eps}
\hspace{10mm}
\includegraphics[width=.45\textwidth]{pttevD0plot.eps}
\caption{Top-quark aN$^3$LO $p_T$ distributions at the LHC (left) and at the Tevatron compared to D0 data \cite{D0pty} (right).}
\label{figptlhc}
\end{center}
\end{figure}

In the left plot of Fig. \ref{figptlhc} we show the aN$^3$LO top-quark $p_T$ distributions \cite{NKpty}, $d\sigma/dp_T$, at 13 and 14 TeV LHC energies. In the right plot of Fig. \ref{figptlhc} we show the aN$^3$LO top-quark $p_T$ distributions \cite{NKpty} at 1.96 TeV Tevatron energy and compare with D0 data \cite{D0pty}, finding very good agreement.

\begin{figure}
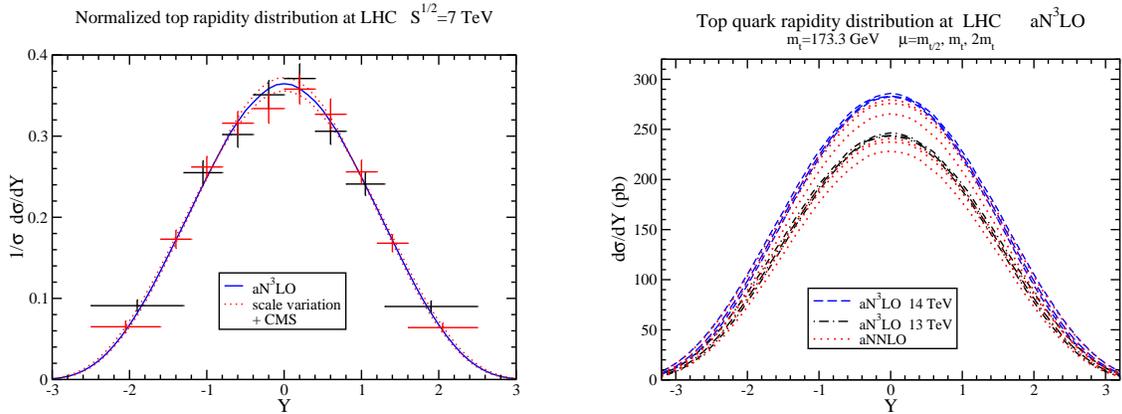

\begin{center}
\includegraphics[width=.45\textwidth]{y7lhcnormCMSplot.eps}
\hspace{10mm}
\includegraphics[width=.45\textwidth]{y13and14lhcaN3LOplot.eps}
\caption{(Left) Top-quark aN$^3$LO normalized rapidity distributions at the 7 TeV LHC and comparison with CMS data \cite{CMStoppty7lhc} in the dilepton (black) and lepton+jets (red) channels; (Right) Top-quark aN$^3$LO rapidity distributions at 13 and 14 TeV LHC energies.}
\label{figylhc}
\end{center}
\end{figure}

We continue with the top-quark rapidity distribution at the LHC \cite{NKpty}. 
In the left plot of Fig. \ref{figylhc} we show the normalized aN$^3$LO top-quark rapidity distribution,  $(1/\sigma) d\sigma/dY$, at 7 TeV LHC energy and compare with results from CMS in the dilepton and lepton+jets channels \cite{CMStoppty7lhc}, finding excellent agreement between theory and data.  The theoretical predictions at 8 TeV are also in excellent agreement with recent CMS top rapidity data in both channels \cite{CMStoppty8lhc}. 
We also show the aN$^3$LO top-quark rapidity distributions, $d\sigma/dY$, at 13 and 14 TeV LHC energies in the right plot of  Fig. \ref{figylhc}.

In the left plot of Fig. \ref{figytev} we compare the  aN$^3$LO distribution of the absolute value of the top-quark rapidity, $d\sigma/d|Y|$, at the Tevatron with D0 data \cite{D0pty} and find very good agreement.

\section{Top-quark forward-backward asymmetry at the Tevatron}

Finally, we discuss the top forward-backward asymmetry at the Tevatron
\beq
A_{\rm FB} =\frac{\sigma(y_t>0)-\sigma(y_t<0)}{\sigma(y_t>0)+\sigma(y_t<0)} \, .
\label{AFB}
\eeq
The above expression can be evaluated with numerator and denominator separately at fixed-order or it can be re-expanded in $\alpha_s$ (see \cite{NKafb} for details through aN$^3$LO). As was discussed in \cite{NKafb} the soft-gluon corrections are dominant and in our formalism they precisely predicted \cite{NKprd84} the exact asymmetry at NNLO.
The high-order perturbative corrections are large: the aN$^3$LO/NNLO ratio is 1.08 without re-expansion in $\alpha_s$, or 1.05 with re-expansion in $\alpha_s$. Including electroweak corrections and the aN$^3$LO QCD corrections we find an asymmetry of ($10.0 \pm 0.6$)\% in the $t{\bar t}$ frame using re-expansion in $\alpha_s$. 

The differential top forward-backward asymmetry is defined by 
\beqa
A^{\rm bin}_{\rm FB} &=&\frac{\sigma^+_{\rm bin}(\Delta y)-\sigma^-_{\rm bin}(\Delta y)}
{\sigma^+_{\rm bin}(\Delta y)+\sigma^-_{\rm bin}(\Delta y)} \hspace{5mm}
{\rm with} \hspace{3mm} \Delta y=y_t-y_{\bar t} \, .
\label{AFBbin}
\nonumber
\eeqa

\begin{figure}
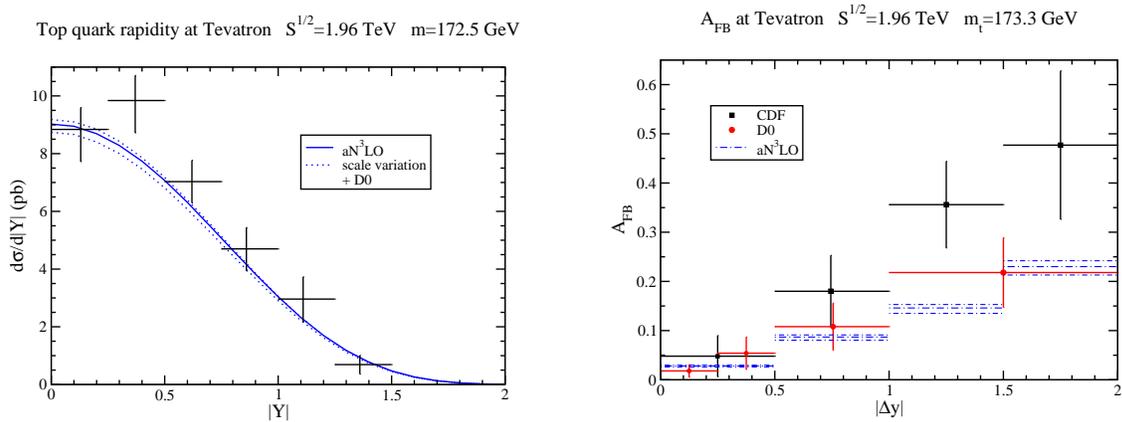

\begin{center}
\includegraphics[width=.45\textwidth]{yabstevD0plot.eps}
\hspace{10mm}
\includegraphics[width=.45\textwidth]{AFBexpplot.eps}
\caption{(Left) Top-quark aN$^3$LO $d\sigma/d|Y|$  distribution at the Tevatron compared with D0 data \cite{D0pty}; (Right) Top-quark aN$^3$LO differential $A_{\rm FB}$ at the Tevatron compared with CDF \cite{CDFafb} and D0 \cite{D0afb} data.}
\label{figytev}
\end{center}
\end{figure}

In the right plot of Fig. \ref{figytev} we plot the differential $A_{\rm FB}$ and compare with recent results from CDF \cite{CDFafb} and D0 \cite{D0afb}. The agreement between theory and experiment is very good for both the total and the differential asymmetries.

\section{Summary}

The N$^3$LO soft-gluon corrections for top-antitop pair production are 
significant and provide the best available theoretical predictions.
Results have been presented for the total $t{\bar t}$ cross sections,
the top-quark $p_T$ and rapidity distributions, and the 
top-quark forward-backward asymmetry.
The corrections are large at LHC and Tevatron energies and they reduce the theoretical uncertainties from scale variation.
There is excellent agreement between aN$^3$LO theoretical predictions and LHC and Tevatron data.

\end{document}